\documentclass[amsmath, amssymb,10pt,aps,prb,twocolumn,notitlepage,showpacs,superscriptaddress]{revtex4-1}
\usepackage{graphicx}
\usepackage{calc}
\usepackage{braket}
\usepackage{ulem}
\usepackage{slashed}
\usepackage{wasysym}
\usepackage{siunitx}
\usepackage{dsfont}
\usepackage{amsthm,amsmath,amsfonts,amssymb,verbatim,color}
\usepackage{amsmath}
\usepackage{graphicx}
\usepackage{subfigure}
\usepackage{bm}
\usepackage{epsfig,slashed}
\usepackage{xcolor}
\usepackage[T1]{fontenc}
\usepackage[colorlinks=true,citecolor=blue,linkcolor=blue,urlcolor=blue]{hyperref}
\usepackage[version=4]{mhchem}

\begin{document}


\title{Observation of magnetically switchable quantum geometric photocurrents}

\author{Qi Tian}
\affiliation{Department of Physics and Astronomy, University of Pennsylvania, Philadelphia, Pennsylvania 19104, USA}
\author{Zhuoliang Ni}
\affiliation{Department of Physics and Astronomy, University of Pennsylvania, Philadelphia, Pennsylvania 19104, USA}
\author{Matthew Cothrine}
\affiliation{Department of Materials Science and Engineering, University of Tennessee, Knoxville, TN 37996, U.S.A.}
\author{David G. Mandrus}
\affiliation{Department of Materials Science and Engineering, University of Tennessee, Knoxville, TN 37996, U.S.A.}
\author{Eugene J. Mele}
\affiliation{Department of Physics and Astronomy, University of Pennsylvania, Philadelphia, Pennsylvania 19104, USA}
\author{Andrew M. Rappe}
\affiliation{Department of Chemistry, University of Pennsylvania, Philadelphia, Pennsylvania 19104, USA}
\author{Charles L. Kane}
\affiliation{Department of Physics and Astronomy, University of Pennsylvania, Philadelphia, Pennsylvania 19104, USA}
\author{Fernando de Juan}
\affiliation{Donostia International Physics Center, Donostia-San Sebastián 20018, Spain}
\affiliation{Ikerbasque, Basque Foundation for Science, Plaza Euskadi 5, 48009 Bilbao, Spain}
\author{Liang Wu}
\email{liangwu@sas.upenn.edu}
\affiliation{Department of Physics and Astronomy, University of Pennsylvania, Philadelphia, Pennsylvania 19104, USA}

\begin{abstract}
In non-centrosymmetric materials, light can be rectified into two types of DC photocurrents, known as injection and shift currents, through the bulk photovoltaic effect. Recent theory has uncovered their deep relation with the two-state quantum geometry of resonant transitions: In non-magnetic crystals, where these currents have been routinely observed, the injection current responds to circular light and probes the Berry curvature, while the shift current responds to linear light and probes the geometric connection. Magnetic crystals have been predicted to show a new set of hitherto unobserved magnetically switchable photocurrents, with the roles of linear and circular light interchanged: A linear injection current, which probes the quantum metric, and a circular shift current, which probes the geometric torsion. In this work, we demonstrate the existence of such currents  for the first time, demonstrating the switching of the current by flipping the Néel vector in a van der Waals antiferromagnet. Furthermore,  their specific frequency and temperature dependence confirm the assignment of circular shift and linear injection currents. Our work demonstrates a new way to control photocurrents in magnets that are directly tied to geometry and have promising applications in antiferromagnetic spintronics and light harvesting.
\end{abstract}

\pacs{}
\maketitle


\noindent\textbf{Introduction}

Quantum geometry has emerged as a promising new principle 
to describe and quantify the quantum features of a broad set of phenomena in materials\cite{torma2023essay,yu2025quantum,jiang2025revealing,liu2025quantum,wang2023quantum,gao2023quantum,kang2025measurements}. The basic idea rests on augmenting the behavior of electrons in valence bands with additional mathematical ingredients that encode details of their quantum mechanical wave functions (also known as the \textit{ground-state} quantum geometry). Such quantum-geometric quantities have been shown to play an essential role in diverse physical phenomena, from the transport and optical properties of  materials to superconductivity in flat band systems\cite{torma2023essay,yu2025quantum}, and can be exploited to predict and design new materials with novel functionalities\cite{Ahn22,jiang2025revealing}. The Berry curvature\cite{xiao2010berry} is perhaps the best known quantum-geometric property, which leads to various robust phenomena such as the quantum Hall effect and the quantized circular photo-galvanic effect\cite{de2017quantized}. Beyond ground state properties, further quantum geometric structure emerges from excitations between states, such as \textit{two-state} geometric quantities\cite{Ahn22,RN15,jiang2025revealing}. In the realm of nonlinear optics, such properties are largely understood for nonmagnetic semiconductors and band insulators, while comparable signatures of \textit {two-state} quantum geometry in the electromagnetic responses of quantum magnets remain a key open challenge to demonstrate experimentally despite their recent theoretical predictions on new geometric quantities\cite{Ahn22,RN15}. 

\begin{figure*}
\centering
\includegraphics[width=0.9\textwidth]{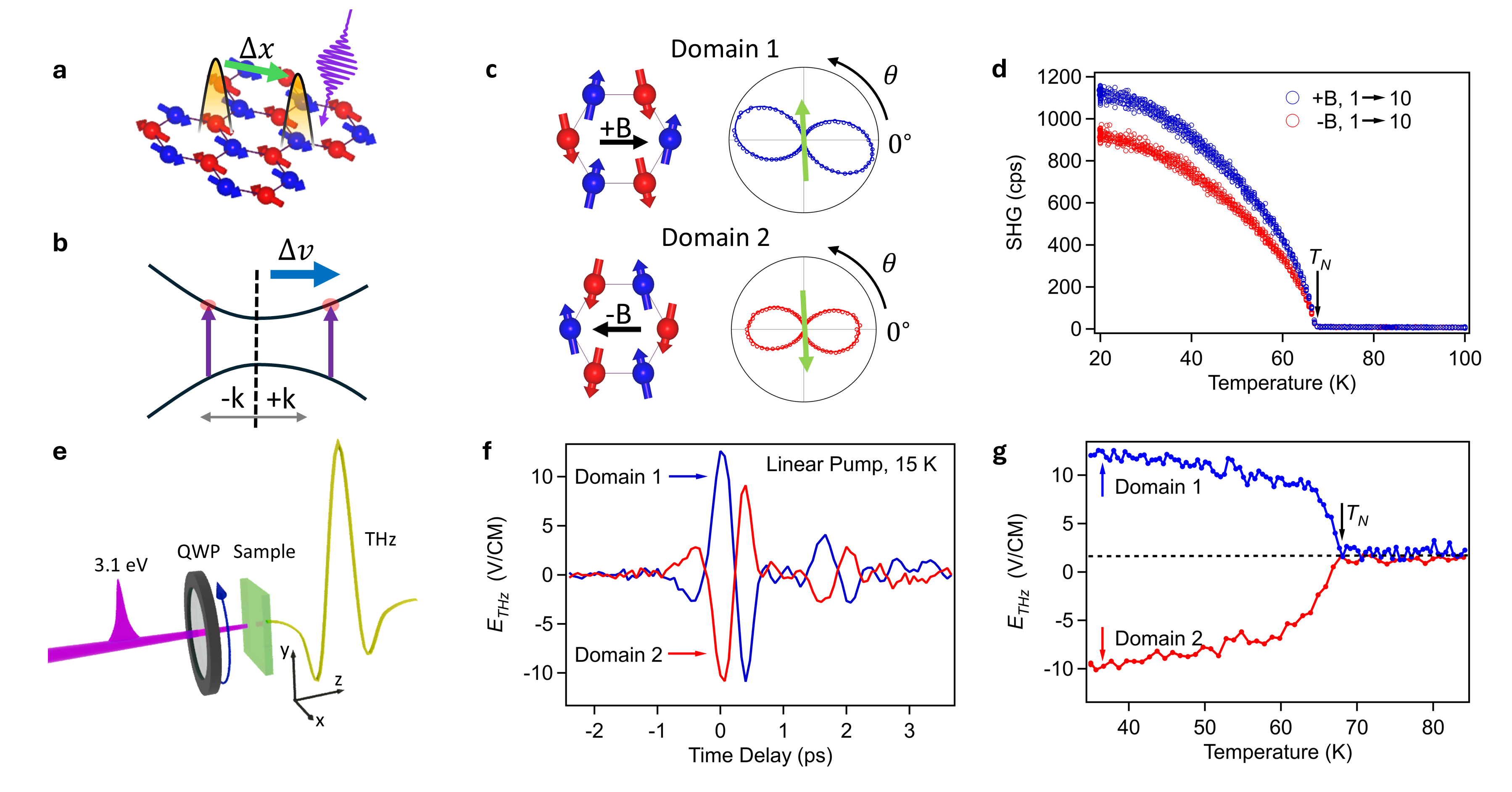}
\caption{\textbf{Observation of the switchable linear injection current.}
(a) Illustration of magnetic shift current. The red and blue arrows represent Mn atoms with opposite spin directions. The two light yellow regions indicate the electron density. Upon optical excitation, a shift of electron density occurs in real space, generating a net current.
(b) Illustration of magnetic injection current: electrons excited at \(\pm \mathbf{k}\) momentum acquire different velocities due to asymmetric band structures across the Brillouin zone. 
(c) Left panels: Two AFM domains in MnPSe\textsubscript{3} with a small canting induced by the magnetic field. The symbols ±B indicate an external in-plane magnetic field perpendicular to the Néel vector. Right panels: cross SHG polar patterns at 20~K from the two domains. Dots and solid lines  represent the measurement data and best fit respectively. The green arrow  indicates the direction of the Néel vector direction from fitting. 
(d) The SHG from the two domains near the peak in (c) as a function of temperature.  Field cool 10 times under each magnetic field direction.  
(e) Diagram of the THz emission setup. 
A 3.1~eV laser induces currents in the sample in the xy plane, which  emits THz pulses propagating in the z-direction.   A QWP is placed before the sample to control polarization. 
(f) Measured THz electric field by the linearly polarized pump pulse from the two domains at 15 K. 
(g) Peak THz electric fields in (f) at a fixed time delay (0 ps) from the two domains as a function of temperature.}   
\label{Fig1}
\end{figure*}

When electrons and holes are optically excited by light in a crystal, they can acquire both a real space shift and a velocity imbalance during interband transitions~\cite{RN24,RN255}. If the crystal lacks inversion symmetry, the forward and backward electron movements do not cancel, leading to a net current. 
This current generation process, mediated by optical nonlinearities, is known as the bulk photovoltaic effect (BPVE)~\cite{RN22}. The accumulation of real space shifts leads to a DC current which is linear in light intensity $J_{sh}\sim I$,  known as shift current\cite{RN24,RN255,RN22, RN8, PhysRevLett.109.116601} as illustrated in Figs. \ref{Fig1}a,  which is very robust and independent of scattering\cite{Hatada20}. The velocity imbalance rather leads to a current injection as illustrated in Fig. \ref{Fig1}b (i.e. a rate of current growth) $\partial_t J_{in} \sim I$,  which saturates to a constant current at long times due to scattering~\cite{PhysRevB.100.195305, RN77}. 
Both shift and injection currents have been observed in many non-magnetic, noncentrosymmetric semiconductors \cite{RN9, RN7, RN10, RN77,hatada2020defect} and more recently in topological semimetals~\cite{rees2020helicity, ninpj2020,maNatPhys2017,maNatMat2019,jiNatMat2019,gaoNatComm2020,siricaPRL2019,osterhoudtNatMat2019}, highlighting a close connection to quantum geometric properties~\cite{RN15, RN21, PhysRevX.10.041041, PhysRevB.95.041104,Ahn22,morimoto2016topological,wu2017giant}.

While the breaking of spatial inversion symmetry is essential for the BPVE, noncentrosymmetry does not necessarily arise from crystal structure alone. In certain antiferromagnets with centrosymmetric lattice structure, the ordered magnetic moments can break both spatial inversion symmetry (\(\mathcal{P}\)) and time-reversal symmetry (\(\mathcal{T}\)), while preserving their combination (\(\mathcal{PT}\))~\cite{RN5, RN18, RN19, RN177, xue2023valley}. Recent theoretical studies have shown that photocurrents can emerge in this setting as well, with two key features that sharply distinguish them from their non-magnetic counterparts. First, these photocurrents are magnetically switchable, i.e., they flow in the opposite direction if the magnetic order is reversed. Second, magnetic shift currents respond only to circular light, while magnetic injection currents respond only to linear light. We name them \textit{circular shift} and \textit{linear injection} currents, respectively.  Thus, at low temperatures where lifetimes are longer, the linear injection effect dominates, which is appealing for applications as it can survive polarization averaging. In addition, both currents have a deep mathematical interpretation in terms of the geometry of interband transitions \cite{PhysRevX.10.041041,Ahn22,Avdoshkin24} (or two-state geometry), which is in general different from ground state quantum geometry. In particular, the linear injection is given by the two-state metric averaged over resonant transitions, while the circular shift is obtained from the torsion of the two-state geometric connection. Interestingly, the frequency integral of the circular shift can be quantized to a universal number in certain topological magnets~\cite{JS24}. Unfortunately, an experimental demonstration of such magnetically switchable photocurrents has remained elusive to date. 


In this work, we present the first experimental observation of linear injection and circular shift currents in the \(\mathcal{PT}\)-invariant antiferromagnet MnPSe\textsubscript{3} using ultrafast THz emission spectroscopy, demonstrating their two unique features. 
First, both photocurrents reverse direction when the antiferromagnetic order is reversed, revealing their magnetic origin. Second, the quotient of the Fourier transformed THz emitted fields scales linearly in frequency, as corresponds to $J_{sh}/\partial_t J_{in}$. Our work thus demonstrates that ultrafast THz photocurrent measurements can be used to trace the different geometric origins of shift and injection currents in magnets, and represents a new method for the electric read-out of the 180 degree antiferromagnetic state switching.\\

\noindent\textbf{Aligning the 180$^{\circ}$ antiferromagnetic domains by a magnetic field}
 
 As a prerequisite to observe magnetically switchable photocurrents, a material is needed where the antiferromagnetic order can be controlled by external means, which is non-trivial in a collinear antiferromagnet. MnPSe\textsubscript{3} belongs to the family of transition metal phosphorus trichalcogenides MPX\textsubscript{3} (M = Mn, Ni, Fe, Co; X = S, Se) and forms a Néel-type AFM order~\cite{RN88, RN12} below $T_N$ = 68 K. (See Fig. \ref{Fig1}a.) The lattice point group is \(\bar{3} (C_{3i})\), which includes an inversion center, thereby prohibiting the BPVE above $T_N$. Below \(T_N\), an in-plane Néel AFM order is established, which transforms as the $E_{u}$ irrep of $\bar{3}$, breaking inversion symmetry and \(C_3\) rotation symmetry, while preserving \(\mathcal{PT}\)  symmetry~\cite{RN88}. As a result, both the circular shift and the linear injection currents are allowed in the magnetic state. MnPSe$_{3}$ contrasts with compounds like bilayer CrI\textsubscript{3}~\cite{RN5} and even-layer MnBi\textsubscript{2}Te\textsubscript{4}~\cite{RN18, RN19}, where the \(C_3\) rotational symmetry forces the circular shift current to vanish. 


The Néel vector should give rise to six possible antiferromagnetic domains oriented at 60$^{\circ}$ from each other. Previous studies have shown that under zero magnetic cooling, the sample goes only into two such domains 180$^{\circ}$ apart under a small uniaxial strain, which can be identified by second harmonic generation (SHG)~\cite{RN12}. We also observe such a phenomenon in a bulk crystal as shown in extended data Fig.~\ref{SHG Field Cooling}. Surprisingly, here we discover that an in-plane magnetic field can select one of the 180$^{\circ}$ antiferromagnetic domains in MnPSe$_3$. Applying an in-plane magnetic field of \(\mathbf{B} = \pm 20\)~mT along the [1-10] direction which is perpendicular to the Néel vector \(\mathbf{L}\), we observe that sweeping the temperature between 20~K and 100~K ten times consistently selects the sample into a single domain (Fig.~\ref{Fig1}c,d), which shows domain switching 180$^{\circ}$ upon reversal of the magnetic field during cooling across $T_N$. Since the magnetic field is inversion even but the Néel vector is inversion odd, the deterministic switching can be explained by the weak breaking of inversion symmetry due to the substrate, which can cause a Dzyaloshinskii–Moriya interaction (see Methods). \\

\noindent\textbf{Magnetically tunable Linear Injection and Circular Shift Currents}

Direct electrical current measurement in MnPSe\textsubscript{3} is not possible because it is an insulator with a gap of 2.27 eV~\cite{grasso1999optical}, with resistivity exceeding \(10^9 \: \Omega\cdot\text{cm}\)~\cite{RN115}. Instead, we detect the emitted terahertz (THz) radiation generated by the ultrafast transient photocurrents, where the radiated field in the far-field limit satisfies \(\mathbf{E}(t) \propto \partial \mathbf{J}(t) / \partial t\) (Fig.~\ref{Fig1}e). We first apply linearly polarized pump pulses centered at 400 nm and measure the THz emission from an approximately 100~$\mu$m-thick MnPSe$_3$ flake at 15~K in the two field-cooled domains (see Methods for more experimental details). Fig.~\ref{Fig1}f shows that the resulting THz emission changes sign when the Néel vector flips by 180$^{\circ}$, consistent with a linear injection current. Additionally, we fix the linearly polarized pump and the time delay at the THz peak position and ramp the temperature from 85 K to 15 K under field cooling. Fig.~\ref{Fig1}g demonstrates an antiferromagnetic phase transition around 68 K in the two domains. The good agreement between the onset of the THz signal enhancement and \(T_N\) strongly suggests a magnetic origin. In contrast, the THz electric field in the paramagnetic phase remains temperature-independent, which suggests a surface origin (Fig.~\ref{Lattice}).

\begin{figure*}
\centering
\includegraphics[width=0.9\textwidth]{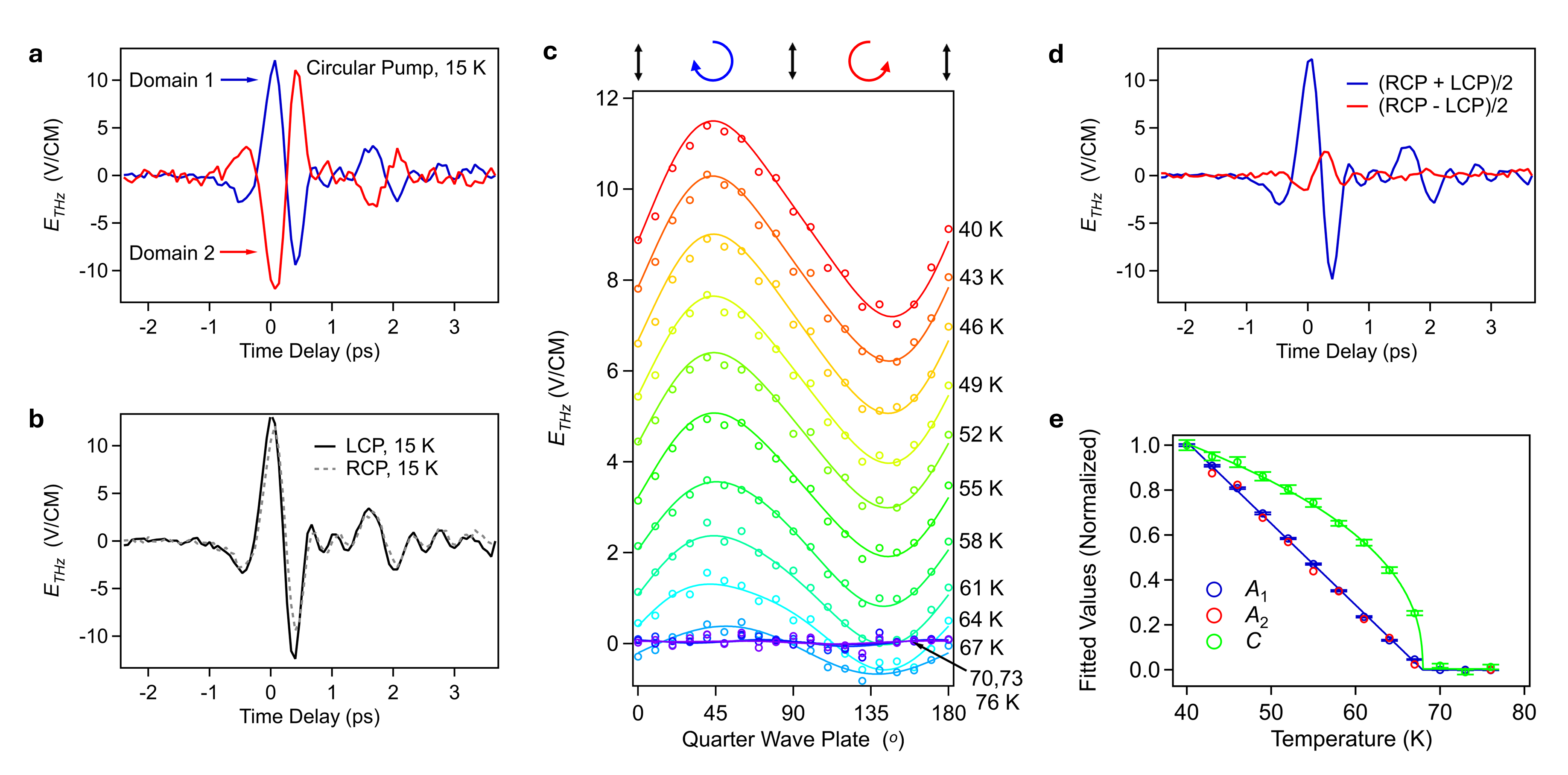}
\caption{
\textbf{Separation of the  circular shift and linear injection currents.} (a) Measured THz electric field under the circularly polarized pump pulse from the two domains at 15 K.
(b) Measured electric field of the THz pulse at 15 K under right and left circularly polarized (RCP and LCP) pump.
(c) Peak THz values at 0 ps as a function of the quarter-wave plate (QWP) angle at a series of temperatures. Open circles represent experimental data, and solid lines represent the fits. 
(d) The helicity-variant and helicity-invariant time traces derived from (b). 
(e) Dots are extracted fitting values from figure (c) and extended data figure S4. Error bars are determined from the fitting uncertainty. Solid lines are fits.
}\label{Fig2}
\end{figure*}

We then apply a circularly polarized pump and measure at 15~K in the two field-cooled domains.  Fig.~\ref{Fig2}a show the resulting THz emission also changes sign when the Néel vector flips by 180$^{\circ}$. The application of circular light can generate both a response to \(|E_x|^2 + |E_y|^2\), which is helicity independent (a linear effect), and a response to \(E_x^*E_y - E_xE_y^*\), which changes sign with helicity and represents the circular photocurrent.  To distinguish these contributions we then apply right- and left-circularly polarized (RCP and LCP) pump pulses at 15~K. The measured THz signals are shown in Fig.~\ref{Fig2}b. To quantify the polarization dependence, we fix the delay stage at the peak position of the THz field under the LCP pump and change the polarization by a quarter waveplate (QWP) while changing the temperature between 40~K and 79~K. The evolution of the resulting THz field is shown in Fig.~\ref{Fig2}c, showing a significant difference between RCP and LCP responses at low temperatures. In contrast, no circularly excited THz signals are observed at 75~K (Fig.~\ref{Lattice}). This absence is attributed to the lattice \(C_3\) rotational symmetry in the paramagnetic state, which forbids a circularly excited photocurrent. Therefore, the circularly excited shift current originates exclusively from the AFM phase. Fig.~\ref{Fig2}d shows the sum and difference of the LCP and RCP responses at 15~K, which correspond to the linear injection and circular shift currents, respectively. In summary, we observe both the circular shift and linear injection currents in the AFM state enabled by the preserved \(\mathcal{PT}\) symmetry in MnPSe\textsubscript{3} \cite{RN255, RN24}. 
\\

\noindent\textbf{Temperature Dependence}

In the previous section, we rely on the \(\mathcal{PT}\) symmetry to assign the shift and injection currents. In the next two sections, we will distinguish them without the need for \(\mathcal{PT}\) symmetry.  Because of the antiferromagnetic origin, the photoconductivity associated with the linear injection and circular shift currents—denoted as \(\eta\) and \(\sigma\), respectively—are both functions of  the order parameter, the Néel vector \(\mathbf{L}\). In the vicinity of \(T_N\), where \(\mathbf{L}\) is small, the linear dependence on \(\mathbf{L}\) dominates over higher-order terms. As a result, we can expand the tensors as \(\sigma^{ijk}(\mathbf{L}) = \sigma^{ijkl}(T > T_N)L^l + O(\mathbf{L}^2)\) (and similarly for \(\eta\)), where \(\sigma^{ijkl}(T > T_N)\) obeys the lattice point group symmetry constraints~\cite{RN20, RN12}. Here we consider the monolayer point group $D_{3d}$ because the photocurrent predominantly flows within individual monolayers~\cite{RN111, xue2023valley}. In the bulk, interlayer stacking breaks mirror symmetry, lowering $D_{3d} \rightarrow C_{3i}$. However, the mirror-asymmetric contributions are expected to be small because the individual monolayer breaks inversion, which is confirmed by our data analysis below. The resulting photocurrents in a single monolayer are thus expressed as:
\begin{align}
    \begin{bmatrix}
        J^I_x \\
        J^I_y
    \end{bmatrix} =& \left[\eta^{\prime}(|E_x|^2 + |E_y|^2)  \label{injection current} + \eta^{\prime\prime}(|E_x|^2 - |E_y|^2) \right]\begin{bmatrix}
        -L_y \\
        L_x
    \end{bmatrix}\notag\\
    & + \eta^{\prime\prime}(E_x^*E_y + E_xE_y^*)\begin{bmatrix}
        L_x \\
        L_y
    \end{bmatrix}\\
     \begin{bmatrix}
        J^S_x \\
        J^S_y
    \end{bmatrix} = &\sigma(E_x^*E_y - E_xE_y^*)\begin{bmatrix}
        L_x \\
        L_y
    \end{bmatrix} \label{Shift current}
\end{align}

The coefficients \(\eta^{\prime} = \frac12(\eta^{xxxy} + \eta^{xyyy})\) and \(\eta^{\prime\prime} = \frac12(\eta^{xxyx} - \eta^{xyxx})\) represent the symmetry-allowed second-order optical conductivities for the injection current, while \(\sigma = \frac12(\sigma^{xxyx} + \sigma^{xyxx})\) is the symmetry-allowed counterpart for the shift current.  See Methods. Equation~(\ref{Shift current}) is consistent with the symmetry-imposed restriction that the shift current can only be excited by circularly polarized light. Moreover, it shows that the shift current always propagates parallel to \(\mathbf{L}\).  In contrast, Eq.~(\ref{injection current}) reveals that the injection current comprises two components: one that always propagates perpendicular to \(\mathbf{L}\), and another that depends on the relative orientation between \(\mathbf{L}\) and the pump polarization. Note the injection current can also be excited by circularly polarized light through the \(|E_x|^2 + |E_y|^2\) term, which which is consistent with what we observed in Fig.~\ref{Fig2}d (blue curve). In fact, Fig.~\ref{Fig2}b shows the dominating signal under circularly polarized light is the injection current. 

In Fig.~\ref{Fig2}c, we employ a quarter-wave plate (QWP) to modulate the pump polarization while continuously measuring the emitted THz field \(E_x(t)\). Linearly and circularly polarized excitations result in distinct angular dependencies, specifically:
\begin{align}
    E_x(t) \propto A_1(t) + A_2(t)\cos{(4\phi - \phi_0)} + C(t)\sin{2\phi} \label{QWP}
\end{align}
where \(\phi\) denotes the QWP angle, with \(0^\circ\) aligned along the crystal \(x\)-axis. \(\phi_0 = \theta_L\) denotes the Néel vector direction (see Equation S20).
Equation~(\ref{QWP}) shows that linearly excited THz emissions, represented by \(A_1(t)\) and \(A_2(t)\), exhibit either no polarization dependence or a \(\cos{4\phi}\) angular dependence, while the circularly excited THz component \(C(t)\) contributes a \(\sin{2\phi}\) term. This clear separation in polarization dependence allows us to distinguish between the two photocurrent contributions. 

We fix the time delay at 0 ps and measure the polarization dependence from 40 K to 79 K in 3 K increments. To isolate the temperature-dependent photocurrent, we subtract the 79 K data—where no circular dependence is observed—to remove the background THz emission originating from the lattice (surface) (see Fig.~\ref{Fig2}c). The near-zero values observed between 70 K and 76 K confirm that the lattice emission is temperature-independent, consistent with the results in Fig.~\ref{Fig1}g. The data are then fitted using Eq.~(\ref{QWP}), and the extracted parameters are provided in Extended Data Fig.~\ref{HWPandQWP}.
To further confirm the results, we replace the QWP with a half-wave plate (HWP) to maintain linear polarization and repeat the measurement. The results, also shown in Fig.~\ref{HWPandQWP}, show the same trend as the QWP results.

We then plot \(A_1(T)\), \(A_2(T)\), and \(C(T)\) in Fig.~\ref{Fig2}e. Comparing with Eqs.~\ref{injection current} and \ref{Shift current}, we associate \(A_1(T)\) and \(A_2(T)\) with injection currents, while \(C(T)\) corresponds to the shift current (see Equations S20-21). The normalized temperature traces of \(A_1(T)\) and \(A_2(T)\) overlap perfectly, as both arise from injection processes. In contrast, the shift current \(C(T)\) displays a distinct temperature dependence that resembles an order parameter near a phase transition.

Fitting the three components to the form \(\propto (T_N - T)^\beta\), with \(T_N = 68\) K fixed, we find \(C(T) \propto (T_N - T)^{0.41}\), consistent with the critical exponent extracted from second harmonic generation measurements \cite{RN12}. In comparison, the injection current scales as \(A_1(T) \propto (T_N - T)^{1.07}\). This difference can be attributed to the additional temperature dependence of the injection current on the carrier lifetime, as the shift current is expected to be independent on scattering \cite{RN5, RN111}. If the scattering rate of the injection current varies with temperature, the injection and shift currents will display different temperature dependencies. In summary, we demonstrate that the circular shift and linear injection currents have different temperature dependence as the injection current is generally assumed to be dependent on the lifetime in the static limit.\\

\noindent\textbf{Dynamics of Injection and Shift Current}

\begin{figure*}
\centering
\includegraphics[width=0.9\textwidth]{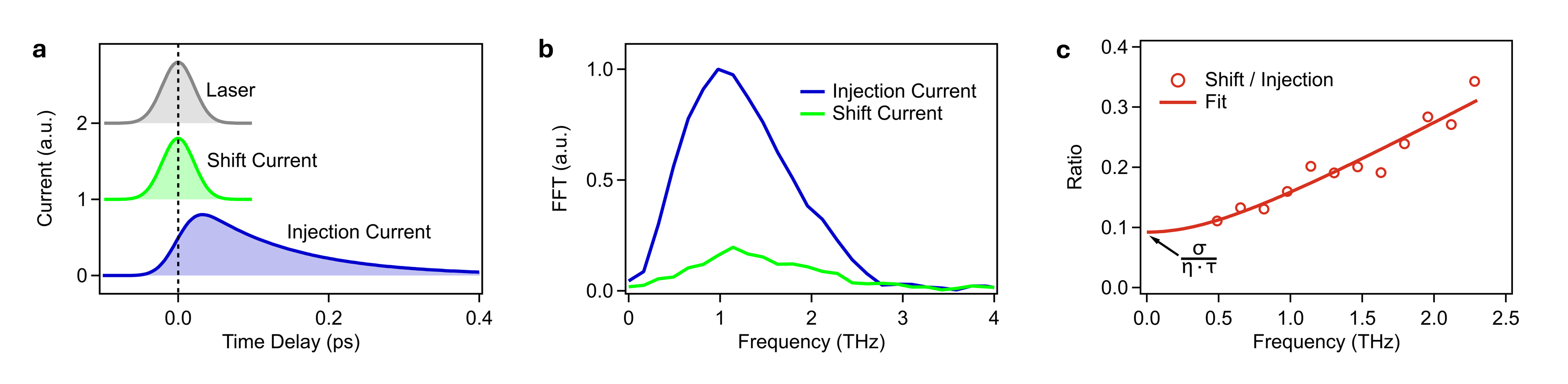}
\caption{
\textbf{Dynamics of the  circular shift and linear injection currents.} (a) Illustrations of the temporal profiles of the pump laser (gray),  the shift current (green), and the injection current (blue).
(b) Fourier transform of the data from Fig.~\ref{Fig2}(d).
(c) Circles are the ratio of the shift-current to injection-current spectrum in (b). The solid line is the best fit with Equation (4).}
\label{Fig3}
\end{figure*}

As discussed in Fig. \ref{Fig1}a,b, the shift current is described by \(J_S(t) \propto I_p(t)\), where \(I_p(t)\) is the temporal profile of the laser intensity, while the relaxation of the injection current can be modeled by adding a phenomenological scattering term \(d_t J \propto I_p(t) - 1/\tau_I J(t)\). See Methods.  These equations lead to different time dependences of the shift and injection currents relative to the laser pulse, which are schematically illustrated in Fig. \ref{Fig3}a. This analysis also predicts that the shift current follows the temporal profile of the laser, but a phase difference exists between the shift and injection currents. The measured THz signals in Fig. \ref{Fig2}d confirms the phase shift.


Figure~\ref{Fig3}b shows the measured THz spectrum by performing Fourier transforms of the time-domain pulses shown in Fig. \ref{Fig2}d. 
As discussed in Methods, fitting the FFT in Fig. \ref{Fig3}b requires careful calibration of the laser pulse duration and spectrum filtering of the mirrors of propagating THz beam. Instead, without requiring laser pulse duration and the calibration of the setup filtering effects, dividing the shift current spectrum by the injection current spectrum in Fig. \ref{Fig3}b yields the ratio of the two photoconductivities, providing a reliable measure of the scattering time of the injection current: 
\begin{align}
    \frac{E_{s}(\Omega)}{E_{i}(\Omega)} 
        = \frac{\sigma}{\eta} (1/\tau_{i} + i\Omega). 
        \label{Ratio of Shift over Injection}
\end{align}
See Methods. 
Fig. \ref{Fig3}c indeed shows a linear scaling of this quotient with frequency for high frequencies, further supporting the identification of shift vs injection photocurrents. Also, the fitting gives rises to an injection current relaxation time of 223 ($\pm48$) fs, which is an order of magnitude larger than that of the topological semimetals\cite{RN123,rees2020helicity,ninpj2020}. This large relaxation time enables the differentiation of shift and injection currents by their time dependence in this work.\\


\noindent\textbf{Discussion}

In summary, we demonstrate the existence of linear injection and circular shift currents in a magnet for the first time. By flipping the Néel vector direction by 180 degrees through in-plane magnetic field cooling perpendicular to the spins, we observe that the propagation directions of these currents also reverse. In collinear antiferromagnets, the 180 degree AFM domain switching has been difficult to induce and read out.  In this work, we demonstrate that photocurrent is a new method for the electric read-out of the 180 degree domain switching. Consequently, we identify two switchable currents, which lie the foundation for advancements in AFM spintronics. Looking forward, we also hope our work generate more interest in using nonlinear optical responses for studying quantum geometry and THz AFM spintronics.

 \section{Acknowledgments}
The project is supported by L.W.'s start-up package at the University of Pennsylvania. The construction of cryogenic terahertz setup  was supported by the Army Research Office under Grant Number W911NF-20-2-0166 and W911NF-25-2-0016.  The SHG work was supported by  the US Office of Naval Research  through the grant N00014-24-1-2064. F. J. acknowledges support from grant PID2021-128760NB0-I00 funded by MCIN/AEI/10.13039/501100011033 and from the 2024 Leonardo Grant for Scientific Research and Cultural Creation from the BBVA Foundation. 

\section{Author Contribution}
Q.T. and N.Z. contribute equally to this work. L.W. conceived and supervised the project. Q. T. and Z.N. performed the experiments and analyzed the data.   M.C. and D.M. grew the crystals.  Q. T., E. M., and A. R., C.K., F. d. J. and L. W. performed the theory analysis.  Q. T., F. d. J. and L. W. wrote the manuscript from input of all authors. All authors edited the manuscript. 

\textit{Data availability:} All data needed to evaluate the conclusions in the paper are present in the paper and the Methods. Additional data related to this paper could be requested from the authors.

\textit {Code availability}:  Codes are not used in this work.


 \textit{Competing Interests: }The authors declare that they have no  competing financial interests.

 \textit{Correspondence: }Correspondence and requests for materials should be addressed to L.W. (liangwu@sas.upenn.edu)

\bibliography{Citation}


\clearpage
\newpage
\widetext  
\setcounter{equation}{0}
\setcounter{figure}{0}
\setcounter{table}{0}
\renewcommand{\theequation}{S\arabic{equation}}
\renewcommand{\thefigure}{S\arabic{figure}}

\newpage
\begin{center}
\textbf{\large Extended data Figure 1}    
\end{center}

\begin{figure*}[h]
\centering
\includegraphics[width=\textwidth]{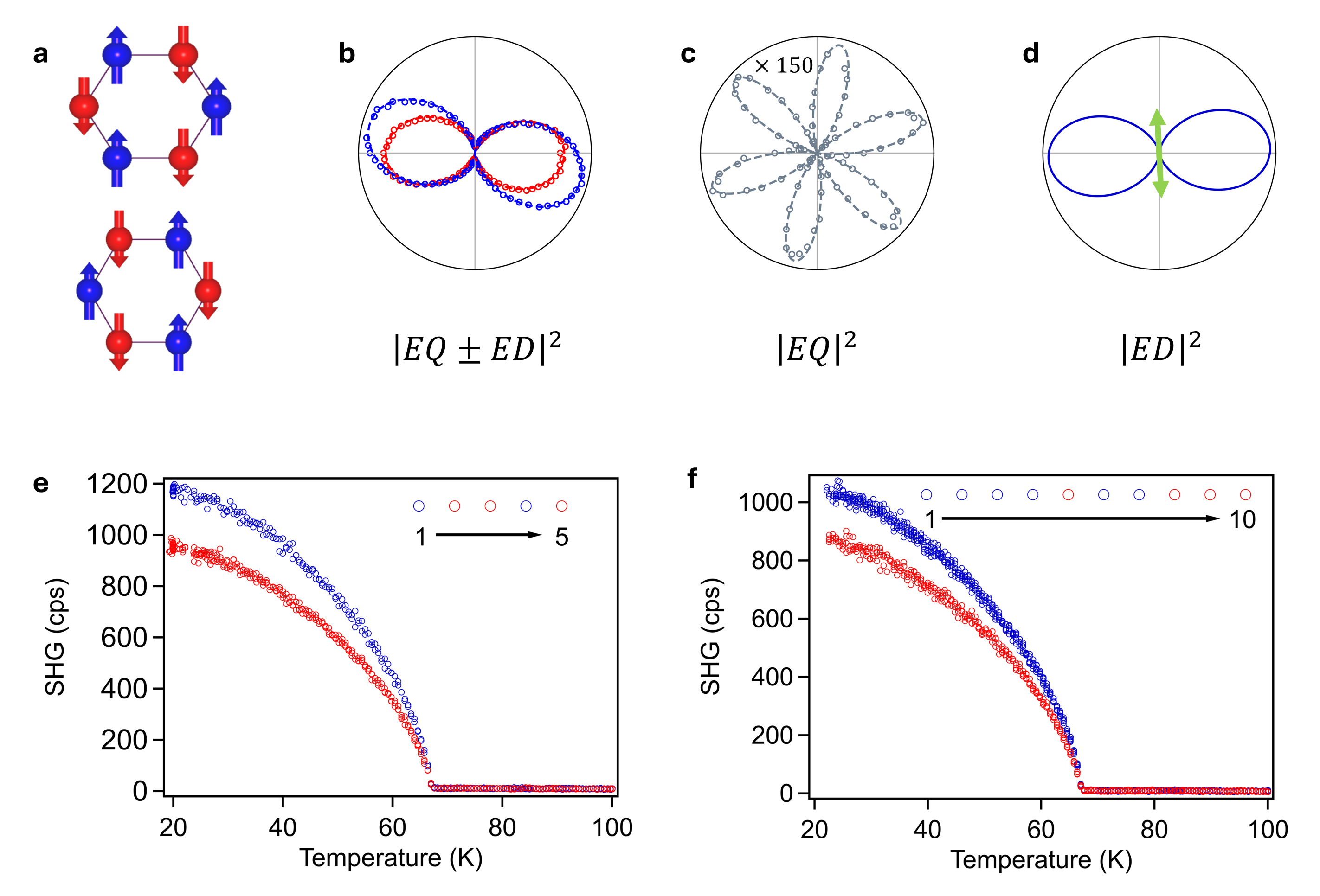}
\caption{\textbf{SHG response under zero-field cooling and field cooling along the spin direction.}
(a) Schematic of the two 180$^\circ$ antiferromagnetic (AFM) domains in MnPSe$_3$ at zero magnetic field. 
(b) SHG polar patterns of the two field-cooled AFM domains measured at 20 K, reproduced from Fig. 1c in the main text. Open circles are the experimental data, and solid lines are fits. 
(c) SHG polar pattern measured at 80 K, above $T_N$. As established previously for MnPSe$_3$~\cite{RN12}, the lattice electric-quadrupole SHG contribution, $E^{EQ}$, is allowed in the centrosymmetric phase and is therefore present at all temperatures. Above $T_N$, the SHG signal is dominated by this contribution and follows the threefold rotational symmetry of the lattice. 
(d) Extracted AFM-induced electric-dipole SHG contribution, $E^{ED}$, after separating the lattice electric-quadrupole background. Below $T_N$, the parity-odd N\'eel order induces an additional electric-dipole SHG contribution. The measured SHG intensity can therefore be described by $I_{\mathrm{SHG}}(\theta)\propto\left|\pm E^{ED}\sin(\theta-\theta_{\mathrm{Neel}})+E^{EQ}\sin 3(\theta-\theta_{\mathrm{lattice}})\right|^2$. Here, $\theta_{\mathrm{Neel}}$ and $\theta_{\mathrm{lattice}}$ denote the N\'eel-vector direction and lattice-axis direction, respectively. In the fitting, $E^{EQ}$ and $\theta_{\mathrm{lattice}}$ are first obtained from the high-temperature data above $T_N$ in panel (c), and are then fixed when analyzing the low-temperature AFM-domain patterns in panel (b). 
(e) Temperature dependence of the SHG intensity under zero-field cooling. The temperature sweep was repeated for five consecutive cycles. 
(f) Temperature dependence of the SHG intensity when the magnetic field was applied along the spin direction during cooling. The temperature sweep was repeated for ten consecutive cycles. In both panels (e) and (f), the system randomly enters one of the two 180$^\circ$ AFM domains.}
\label{SHG Field Cooling}
\end{figure*}



\newpage
\begin{center}
\textbf{\large Extended Data Figure 2}    
\end{center}

\begin{figure*}[h]
\centering
\includegraphics[width=\textwidth]{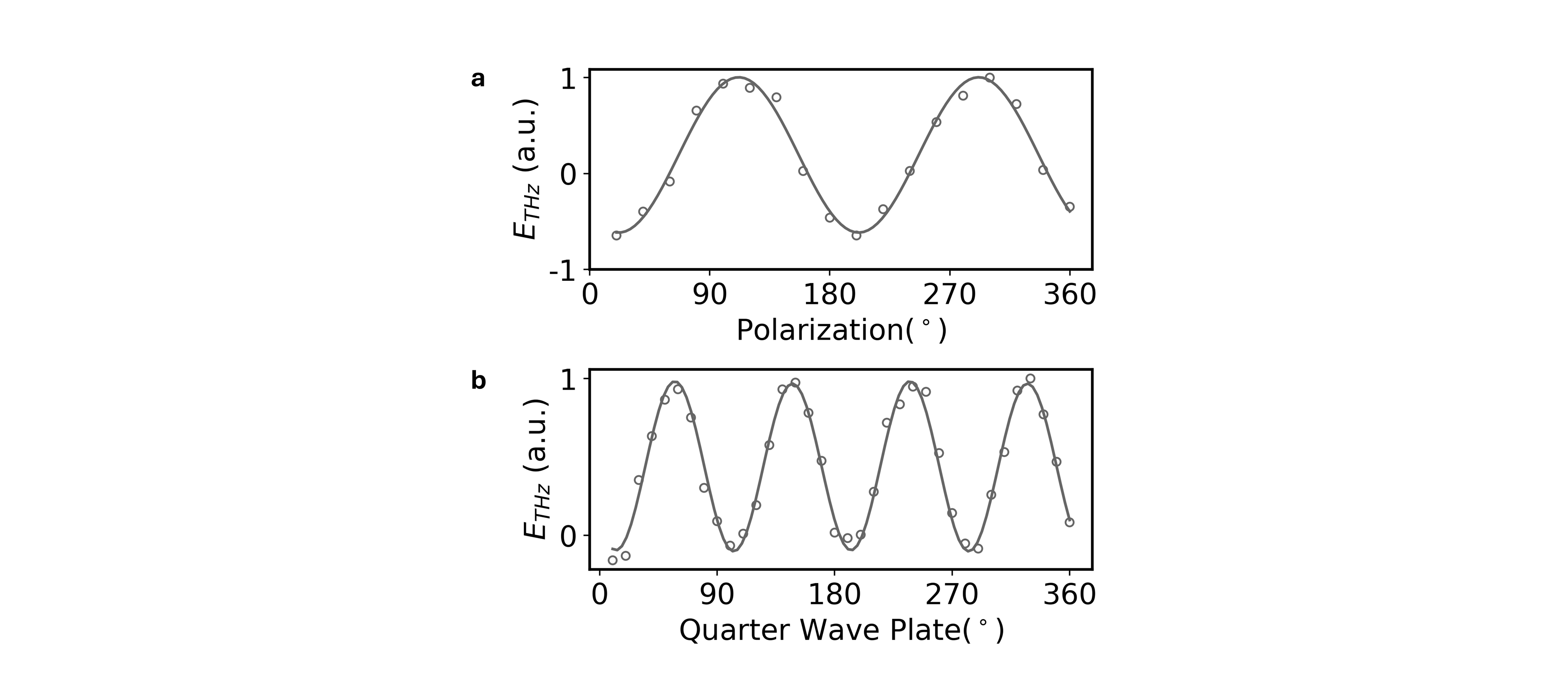}
\caption{\textbf{THz emission above $T_N$.}
Peak values of emitted THz electric field (at 0 ps) at 75 K as a function of pump
polarization. (a) We use a HWP for linear polarization rotation; (b) We uses a QWP
 to periodically alternate between linear and circular polarization. The horizontal axis is the fast axis angle of the QWP. Solid lines represent best fits and the circles are the data. In (a), the data are fitted with \(A\cos{2\theta} + B\sin{2\theta}\), which is derived with the constraint from the lattice \(C_3\) rotation symmetry. \(\theta\) is the pump polarization. To verify the absence of circularly excited currents, we used a QWP with the fast axis angle $\phi$ to vary the pump polarization from linear to circular and vice versa. In this measurement,  contributions of the photocurrent from linearly polarized light should vary as \(\cos{4\phi}\), as the linear component intensity also gets modulated, while any circular component should vary as \(\cos{2\phi}\). The excellent fit to \(\cos{4\phi}\) in (b) confirms the absence of circularly excited photocurrent.}
\label{Lattice}
\end{figure*}

\newpage
\begin{center}
\textbf{\large Extended data Figure 3}    
\end{center}

\begin{figure*}[h]
\centering
\includegraphics[width=\textwidth]{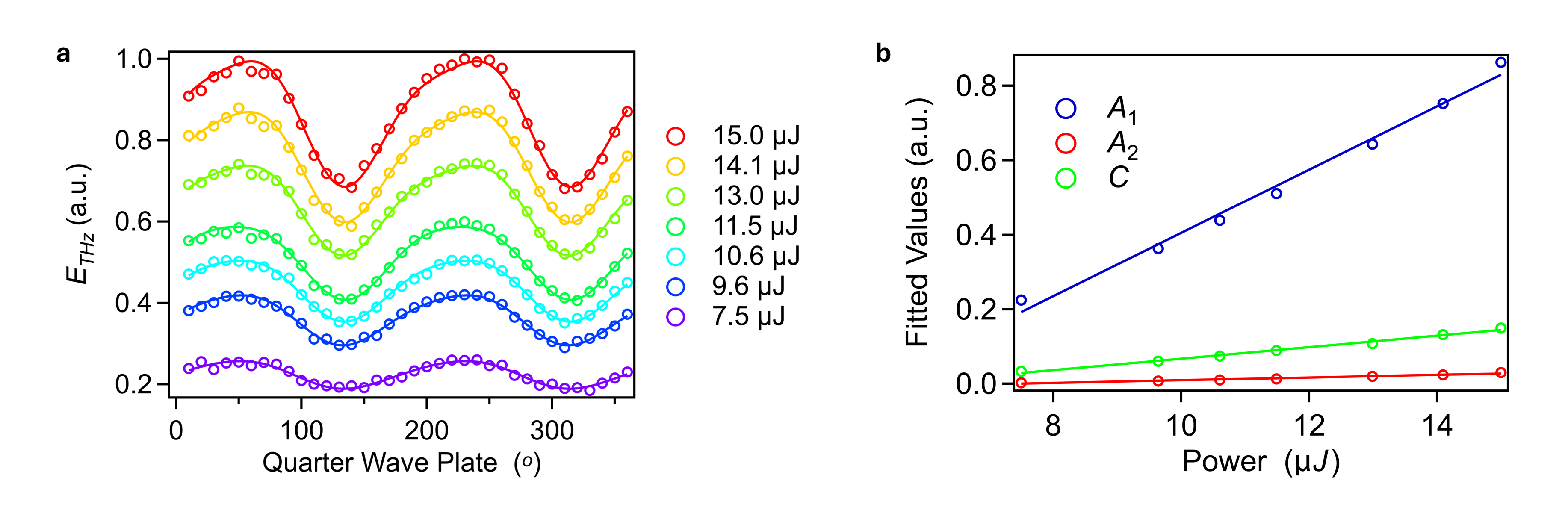}
\caption{\textbf{Power dependence.}
(a) Dependence of the peak values of the emitted THz time trace on the fast axis angle of the quarter-wave plate. The pump size is 1 mm$^2$, and the single pulse energy varies from 7.5 $\mu$J  to 15 $\mu$J. (b) Extracted fitted values. The power dependence is also used to calibrate the heating effect induced by the pump beam.}
\label{Power}
\end{figure*}

\newpage
\begin{center}
\textbf{\large Extended data Figure 4}    
\end{center}

\begin{figure*}[h]
\centering
\includegraphics[width=\textwidth]{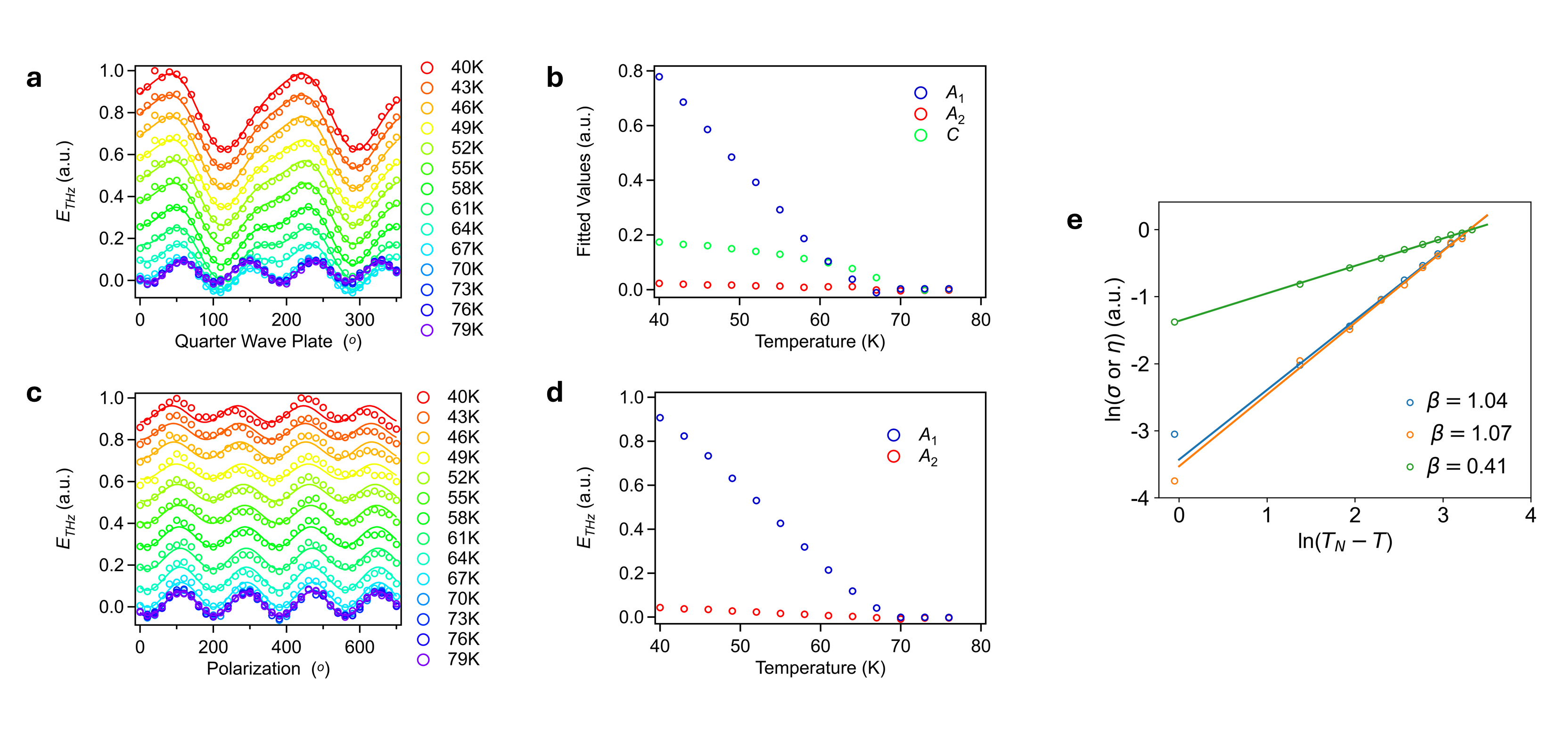}
\caption{\textbf{Temperature dependence.} (a) Peak of the emitted THz time trace as a function of the quarter-wave plate fast axis angle. Open circles represent the measured data, and solid lines are the best fits. (b) Extracted fit parameters in (a), with the values at 79 K subtracted to highlight the magnitude change associated with the AFM order. (c) Peak of the emitted THz time trace as a function of linear pump polarization. Open circles represent the measured data, and solid lines are the fits. (d) Extracted fit parameters in (c), with the values at 79 K subtracted to highlight the  change associated with the AFM order. (e) Fit of $\ln{\sigma \text{ or } \eta}$ versus $\ln{(T_N - T)}$, where $\sigma, \eta^{\prime}, \eta^{\prime\prime}$ are normalized data from panels (b) and (d). The $T_N$ value is fixed at 68 K, which is determined from fitting $\sigma$ directly with $(T_N - T)^\beta$.}
\label{HWPandQWP}
\end{figure*}

\clearpage
\widetext

\section{Methods}
\noindent\textbf{THz and optics setups}\\

\begin{figure*}[h]
\centering
\includegraphics[width=\textwidth]{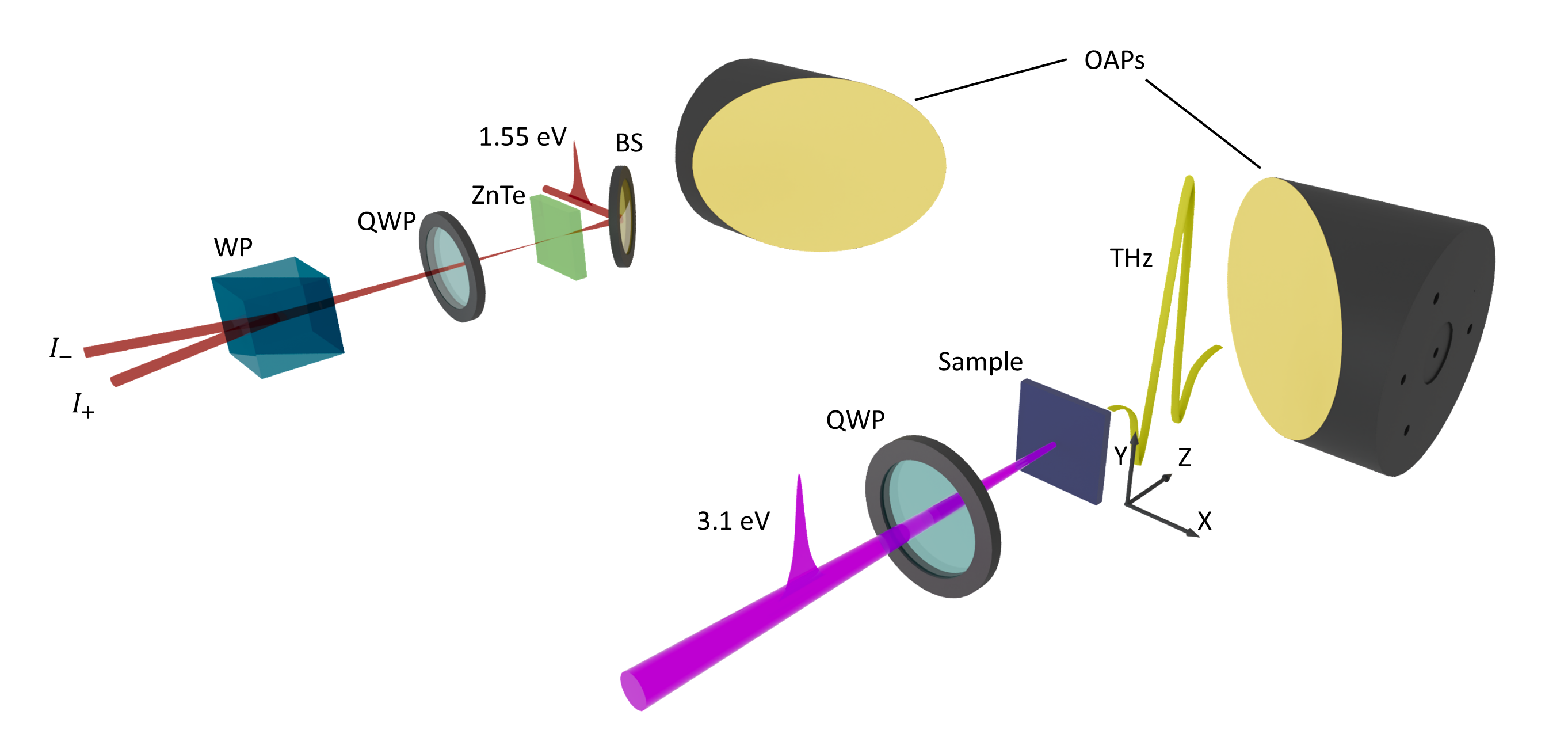}
\caption{\textbf{Schematic of the THz emission setup.} An 800~nm (1.55~eV), around 50~fs, 1~kHz repetition rate laser is spit into two beams by a R:T = 80:20 beam splitter. The reflected laser (80\%) is focused onto the sample using a lens with focal length \(f = 75~\unit{cm}\). A 0.1~mm thick beta-BaB\textsubscript{2}O\textsubscript{4} (BBO) crystal is employed to double the frequency of the 800~nm laser into 400~nm (3.1~eV), while preserving the pulse duration. A quarter-wave plate (QWP) is put before the sample to modulate the laser polarization. 
To collect and focus the emitted THz pulses onto the electro-optic (EO) sampling crystal, a pair of 3-inch diameter, \(f = 6~\unit{inch}\) off-axis parabolic (OAP) mirrors are used. The sample and the EO-sampling crystal are put at the focus of the two OAPs, respectively. For EO sampling, we use a 1~mm thick, \(\left<110\right>\)-cut ZnTe crystal and the attenuated transmitted 800 nm laser. The EO-sampling crystal is set to measure the x-component of emitted THz. The xyz coordinate system represents the lab orientation, with the y-axis oriented vertically and the z-axis along the laser propagation direction. A combination of a QWP, a Wollaston prism (WP), and a balanced photodetector is employed to detect the EO signal. The absolute THz electric field is calibrated from the normalized balanced-detector signal in the EO-sampling setup. In the small-signal limit, the detected EO signal is proportional to the phase retardation induced in the ZnTe crystal by the THz field. The incident THz field is then estimated as
$E_{\mathrm{THz}}=\left[n_{\mathrm{ZnTe}}(\mathrm{THz})+1\right]\lambda\left((I_1 - I_2)/(I_1+I_2)\right)/\left[4\pi n_{\mathrm{ZnTe}}^3(800~\mathrm{nm})r_{41}L\right]$,
where $(I_1 - I_2)/(I_1+I_2)$ is the normalized balanced-detector output, proportional to the THz electric field induced birefringence in ZnTe, $\lambda=800$~nm is the probe wavelength, $n_{\mathrm{ZnTe}}(\mathrm{THz})$ and $n_{\mathrm{ZnTe}}(800~\mathrm{nm})$ are the refractive indices of ZnTe in the THz range and at 800~nm, respectively, $r_{41}$ is the EO coefficient of ZnTe, and $L$ is the effective interaction length. The prefactor $\left[n_{\mathrm{ZnTe}}(\mathrm{THz})+1\right]/2$ accounts for the Fresnel transmission of the THz field from air into the ZnTe crystal. Because the 1-mm-thick ZnTe crystal is used for EO sampling and the bandwidth of interest is below the phonon-resonance-limited regime, we approximate $L$ by the ZnTe thickness.}
\label{Setup}
\end{figure*}


In the SHG measurement, we use an 800 nm, 80 MHz laser with a pulse duration of approximately 50 fs as the fundamental source. The beam is focused to a spot size of $\sim$2~$\mu$m using a 50$\times$ objective lens, which also serves to collimate the reflected 400 nm SHG signal. A half-wave plate and a linear polarizer, both mounted on motorized rotators, are used to measure the SHG polar patterns.
\\

\noindent\textbf{THz emission spectrum from injection and shift current}\\
As discussed in the main text, the injection current \(J_i\) and shift current \(J_s\) are excited by
\(
\frac{dJ_{i}(t)}{dt} = \eta I_p(t), J_s(t) = \sigma I_p(t),
\)
where \(\eta\) and \(\sigma\) denote the photoconductivities associated with the injection and shift currents, respectively. Their units differ by a factor of \([s^{-1}]\), reflecting the distinct excitation mechanisms.  
The temporal profile of the pump laser is assumed to follow a Gaussian form,
\(
I_p(t) = I_0 e^{-t^2/\sigma^2}.
\)
Phenomenologically, we include a scattering term \(\frac{dJ(t)}{dt} \propto -\frac{1}{\tau} J(t)\) for the injection current, yielding
\begin{align}
    \frac{dJ_{i}(t)}{dt} &= \eta I_p(t) - \frac{1}{\tau_{i}} J_{i}(t), 
    \label{dynamics of injection current}
\end{align}
where \(\tau_i\) is the characteristic scattering time of the injection current.  
The solution to Eq.~\eqref{dynamics of injection current} is
\begin{align}
    J_{i}(t) &= \eta \big[\Theta(t)\, e^{-t/\tau_i}\big] * I_p(t), 
    \label{time solution of injection current}
\end{align}
where \(\Theta(t)\) is the Heaviside step function and * denotes convolution in time.  
Taking the Fourier transform of Eq.~\eqref{dynamics of injection current} gives
\begin{align}
    J_{i}(\Omega) &= \eta \frac{1}{1/\tau_{i} + i\Omega}\, I_p(\Omega), 
    \label{spectrum of injection current}
\end{align}
with \(I_p(\Omega) = I_0 e^{-(\sigma/2)^2 \cdot \Omega^2}\), and \(\Omega = 2\pi f\) is the angular frequency. 
For the shift current, the Fourier spectrum follows directly as
\begin{align}
    J_{s}(\Omega) &= \sigma I_p(\Omega).
    \label{spectrum of shift current}
\end{align}

In the far-field approximation, the emitted THz pulse is proportional to the time derivative of the current, \(E_{\text{emit}} \propto \frac{\partial J(t)}{\partial t}\). Substituting the above expressions, we obtain:
\begin{align}
    E_{i}(\Omega) &\propto \eta \frac{i\Omega}{1/\tau_{i} + i\Omega}\, I_p(\Omega), 
    \label{THz spectrum of injection current} \\
    E_{s}(\Omega) &\propto \sigma\, i\Omega\, I_p(\Omega),  
    \label{THz spectrum of shift current}
\end{align}
and consequently,
\begin{align}
    \frac{E_{s}(\Omega)}{E_{i}(\Omega)} 
        &= \frac{\sigma}{\eta} \Big( \frac{1}{\tau_{i}} + i\Omega \Big).
        \label{Ratio of Shift over Injection}
\end{align}

Equations~\eqref{THz spectrum of injection current} and~\eqref{THz spectrum of shift current} describe the frequency-domain THz emission spectra associated with injection and shift currents, respectively.  

In the THz emission setup, one pair of 3-inch diameter, \(f = 6~\unit{inch}\) OAP mirrors are used for the collection and focus of THz pulses. Due to the finite aperture size, low-frequency THz photons experience a lower collection ratio. This introduces a frequency-dependent detection response arising from the OAP filtering effect. Besides, the EO-sampling crystal, ZnTe, exhibits a phonon resonance at approximately 5.3~THz, which causes a diverging refractive index in that vicinity. As a result, the phase matching between the 800~nm probe and the THz pulse becomes poor above 3~THz. This limits our detection bandwidth to below 3~THz and effectively acts as a spectral filtering function. In conclusion, the measured THz spectrum is subject to frequency-dependent filtering effects arising from both THz propagation and EO sampling. Crucially, Eq.~\eqref{Ratio of Shift over Injection} eliminates these filtering effects, as well as the dependence on the pump spectrum, thereby enabling a direct determination of the injection-current lifetime and the photoconductivity ratio of $\sigma/\eta$.
\\

\noindent\textbf{Symmetry calculation with rank 4 tensor in MnPSe\textsubscript{3}}\\
As discussed in the main text, the magnetic photocurrents are described by the expression  
\begin{align}
    J_i = \sigma^{ijkl}(0; -\omega, \omega)\, E_j(-\omega)\, E_k(\omega)\, L_l + O(\mathbf{L^3}) \label{Rank4 Equation}
\end{align}
where \(\mathbf{J}\), \(\mathbf{E}\), and \(\mathbf{L}\) represent the photocurrent, electric field, and Néel vector, respectively. The tensor \(\sigma\) represents the general nonlinear conductivity, constrained by the lattice symmetry. The expansion on \(\mathbf{L}^2\) is prohibited by space inversion symmetry. Since measurements are performed under normal incidence and the photocurrent flows primarily in-plane, both \(\mathbf{J}\) and \(\mathbf{E}\) have no out-of-plane components. Likewise, due to the strong XY anisotropy, the Néel vector \(\mathbf{L}\) is also confined to the plane. As a result, we set \(\{i, j, k, l\} = \{x, y\}\), i.e., we calculate with in-plane components only. Expanding equation~\eqref{Rank4 Equation}, we have:
\begin{align}
\left[\begin{array}{c}
J_x \\
J_y
\end{array}\right]_{\text{sym}} &= 
\left[\begin{array}{l}
\left(\sigma_{xxxx} E_x E_x^* + \sigma_{xyyx} E_y E_y^*\right) L_x + \left(\sigma_{xxyy} E_x E_y^* + \sigma_{x y x y} E_y E_x^*\right) L_y \\
\left(\sigma_{y x x y} E_x E_x^* + \sigma_{y y y y} E_y E_y^*\right) L_y + \left(\sigma_{y x y x} E_x E_y^* + \sigma_{y y x x} E_y E_x^*\right) L_x
\end{array}\right] \label{Rank4, symmetric} \\[1em]
\left[\begin{array}{c}
J_x \\
J_y
\end{array}\right]_{\text{asym}} &= 
\left[\begin{array}{l}
\left(\sigma_{x x x y} E_x E_x^* + \sigma_{x y y y} E_y E_y^*\right) L_y + \left(\sigma_{x x y x} E_x E_y^* + \sigma_{x y x x} E_y E_x^*\right) L_x \\
\left(\sigma_{y x x x} E_x E_x^* + \sigma_{y y y x} E_y E_y^*\right) L_x + \left(\sigma_{y x y y} E_x E_y^* + \sigma_{y y x y} E_y E_x^*\right) L_y
\end{array}\right] \label{Rank4, asymmetric}
\end{align}

Here, \(\mathbf{J}_{\text{sym}}\) and \(\mathbf{J}_{\text{asym}}\) respectively collect terms with even and odd numbers of \(x\) indices. With lattice \(C_3\) rotational symmetry, we have the following relations:
\begin{align}
\text{symmetric terms:} \quad 
\begin{cases}
\sigma_{xxxx} = \sigma_{yyyy} \\
\sigma_{xxyy} = \sigma_{yyxx} \\
\sigma_{xyyx} = \sigma_{yxxy} \\
\sigma_{xyxy} = \sigma_{yxyx} \\
\sigma_{xxxx} - \sigma_{xyyx} = \sigma_{xxyy} + \sigma_{xyxy}
\end{cases}
\quad
\text{anti-symmetric terms:} \quad
\begin{cases}
\sigma_{xxxy} = -\sigma_{yyyx} \\
\sigma_{yyxy} = -\sigma_{xxyx} \\
\sigma_{yxyy} = -\sigma_{xyxx} \\
\sigma_{xyyy} = -\sigma_{yxxx} \\
\sigma_{xxxy} - \sigma_{xyyy} = \sigma_{yyxy} + \sigma_{yxyy}
\end{cases} \label{C3 Constraint}
\end{align}

Plugging the constraints from equation~\eqref{C3 Constraint} into equations~\eqref{Rank4, symmetric} and~\eqref{Rank4, asymmetric}, we obtain:
\begin{align}
\left[\begin{array}{c}
J_x \\
J_y
\end{array}\right]_{\text{sym}} 
&= \frac12\left(\sigma_{xxxx} + \sigma_{xyyx}\right)
    \left(\left|E_x\right|^2 + \left|E_y\right|^2\right)
    \left[\begin{array}{c}
    L_x \\
    L_y
    \end{array}\right]
    + i\, \operatorname{Im}(\sigma_{xxyy})
    \left(E_x^* E_y - E_xE_y^*\right)
    \left[\begin{array}{c}
    L_y \\
    -L_x
    \end{array}\right] \notag\\
&\quad + \, \operatorname{Re}(\sigma_{xxyy})
    \left[\begin{array}{c}
    \left(\left|E_x\right|^2 - \left|E_y\right|^2\right) L_x + \left(E_x^* E_y + E_xE_y^*\right) L_y \\
    -\left(\left|E_x\right|^2 - \left|E_y\right|^2\right) L_y + \left(E_x^* E_y + E_xE_y^*\right) L_x
    \end{array}\right] \label{Symmetric Current} \\[1em]
\left[\begin{array}{c}
J_x \\
J_y
\end{array}\right]_{\text{asym}} 
&= \frac12\left(\sigma_{xxxy} + \sigma_{xyyy}\right)
    \left(\left|E_x\right|^2 + \left|E_y\right|^2\right)
    \left[\begin{array}{c}
    L_y \\
    -L_x
    \end{array}\right]
    + i\, \operatorname{Im}(\sigma_{xxyx})
    \left(E_x^* E_y - E_xE_y^* \right)
    \left[\begin{array}{c}
    L_x \\
    L_y
    \end{array}\right] \notag\\
&\quad + \operatorname{Re}(\sigma_{xxyx})
    \left[\begin{array}{c}
    -\left(\left|E_x\right|^2 - \left|E_y\right|^2\right) L_y + \left(E_x^* E_y + E_xE_y^* \right) L_x \\
    -\left(\left|E_x\right|^2 - \left|E_y\right|^2\right) L_x - \left(E_x^* E_y + E_xE_y^* \right) L_y
    \end{array}\right] \label{Asymmetric Current}
\end{align}

It is important to note that we cannot directly apply mirror operations to equation~\eqref{Asymmetric Current} to set \(\sigma_{ijkl}\) with odd numbers of \(x\) to zero, as the Néel vector \(\mathbf{L}\) behaves differently from the polar vectors \(\mathbf{E}\) and \(\mathbf{J}\). In fact, \(\mathbf{L}\) may transform as either an axial or polar vector depending on whether the mirror operation exchanges the two antiferromagnetic sublattices. In the case of MnPSe\textsubscript{3}, we will show later that \(\mathbf{L}\) behaves as axial vector, and the mirror symmetry sets all \(\sigma_{ijkl}\) with even numbers of \(x\) to zero.\\
\\

\noindent\textbf{Symmetry calculation with group invariants during phase transition}\\
We now verify the validity of equations~\eqref{Symmetric Current} and~\eqref{Asymmetric Current} using irreducible representations (irreps) from group theory. In MnPSe\textsubscript{3}, the monolayer belongs to the \(D_{3d}\) point group. However, interlayer stacking breaks mirror symmetry, reducing the symmetry to the \(C_{3i}\) point group. We begin by analyzing the irreps in a monolayer and then examine effect from mirror symmetry breaking.

As discussed in the previous section, only the irreps corresponding to the in-plane components of the involved vectors need to be considered. The character table\cite{aroyo2011crystallography} and irreps of \(D_{3d}\) are summarized in Table~\ref{D3d_and_irrep_tables}. For equation~\eqref{Rank4 Equation} to hold, the tensor product \(\mathbf{E} \otimes \mathbf{E}^* \otimes \mathbf{L}\) must contain the same irrep as \(\mathbf{J}\).

From the equation \(A_{1g} \otimes E_u = E_u\), there is one allowed combination:
\begin{align}
    \begin{bmatrix}
        J_x \\
        J_y
    \end{bmatrix}
    = a(E_x E_x^* + E_y E_y^*)
    \begin{bmatrix}
        -L_y \\
        L_x
    \end{bmatrix}
    \label{irrep_inject1}
\end{align}

From the equation \(A_{2g} \otimes E_u = E_u\), there is one allowed combination as well:
\begin{align}
    \begin{bmatrix}
        J_x \\
        J_y
    \end{bmatrix}
    = c(E_x E_y^* - E_y E_x^*)
    \begin{bmatrix}
        -L_y \\
        L_x
    \end{bmatrix}
    \label{irrep_shift1}
\end{align}

Finally, from the product \(E_u \otimes E_g = A_{1u} \oplus A_{2u} \oplus E_u\), we obtain one additional independent combination:
\begin{align}
    \begin{bmatrix}
        J_x \\
        J_y
    \end{bmatrix}
    = e
    \begin{bmatrix}
        (E_x E_y^* + E_y E_x^*) L_x - (E_x E_x^* - E_y E_y^*) L_y \\
        -(E_x E_x^* - E_y E_y^*) L_x - (E_x E_y^* + E_y E_x^*) L_y
    \end{bmatrix}
    \label{irrep_inject2}
\end{align}

Together, equations~\eqref{irrep_inject1}, \eqref{irrep_shift1}, and~\eqref{irrep_inject2} involve three independent coefficients. Among them, equations~\eqref{irrep_inject1} and~\eqref{irrep_inject2} describe linearly excited processes, while equation~\eqref{irrep_shift1} corresponds to circular excitation. Due to \(\mathcal{PT}\) symmetry, magnetic injection current can only be generated by linearly polarized light, whereas magnetic shift current is induced exclusively by circularly polarized light. Therefore, the magnetic injection current is governed by equations~\eqref{irrep_inject1} and~\eqref{irrep_inject2}, while the magnetic shift current arises solely from equation~\eqref{irrep_shift1}. 


Then we repeat the same analysis using the reduced point group \(C_{3i}\), which lacks mirror symmetry due to layer stacking. There are together six independent combinations. Specifically, 
\begin{align}
    \begin{bmatrix}
        J_x \\
        J_y
    \end{bmatrix}
    =\,
    &\underbrace{
        a (E_x E_x^* + E_y E_y^*) 
        \begin{bmatrix}
            -L_y \\
            L_x
        \end{bmatrix}
        + 
        b (E_x E_x^* + E_y E_y^*)
        \begin{bmatrix}
            L_x \\
            L_y
        \end{bmatrix}
    }_{\text{\(A_{1g} \otimes E_u = E_u\)}} \notag \\
    +\,
    &\underbrace{
        c (E_x E_y^* - E_y E_x^*) 
        \begin{bmatrix}
            -L_y \\
            L_x
        \end{bmatrix}
        + 
        d (E_x E_y^* - E_y E_x^*)
        \begin{bmatrix}
            L_x \\
            L_y
        \end{bmatrix}
    }_{\text{\(A_{2g} \otimes E_u = E_u\)}} \notag \\
    +\,
    &\underbrace{
        e \begin{bmatrix}
            (E_x E_x^* - E_y E_y^*) L_x + (E_x E_y^* + E_y E_x^*) L_y \\
            -(E_x E_y^* + E_y E_x^*) L_x + (E_x E_x^* - E_y E_y^*) L_y
        \end{bmatrix}
        +
        f \begin{bmatrix}
            (E_x E_y^* + E_y E_x^*) L_x - (E_x E_x^* - E_y E_y^*) L_y \\
            -(E_x E_x^* - E_y E_y^*) L_x - (E_x E_y^* + E_y E_x^*) L_y
        \end{bmatrix}
    }_{\text{\(E_u \otimes E_g = A_{1u} \oplus A_{2u} \oplus E_u\)}}
    \label{C3i}
\end{align}

Compared with equations~\eqref{irrep_inject1}, \eqref{irrep_shift1}, and~\eqref{irrep_inject2}, we can easily find that the mirror symmetry sets \(b, d, f = 0\). Besides, comparing equation~\eqref{C3i} with equations~\eqref{Symmetric Current} and~\eqref{Asymmetric Current}, we identify the following correspondences: \(a = \frac12(\sigma_{xxxy} + \sigma_{xyyy})\), \(b = \frac12(\sigma_{xxxx} + \sigma_{xyyx})\), \(c = i\,\mathrm{Im}(\sigma_{xxyx})\), \(d = i\,\mathrm{Im}(\sigma_{xxyy})\), \(e = \mathrm{Re}(\sigma_{xxyx})\), and \(f = \mathrm{Re}(\sigma_{xxyy})\). This confirms the validity of expanding \(\sigma_{ijk}(\mathbf{L}) = \sigma_{ijkl} L_l + O(\mathbf{L^3})\). 

In conclusion, in a monolayer, we have: 
\begin{align}
    \mathbf{J}_{I} =&\eta^{\prime}(|E_x|^2 + |E_y|^2)\mathbf{L} \times \hat{n}  + \eta^{\prime\prime}(|E_x|^2 - |E_y|^2)\mathbf{L} \times \hat{n} 
     + \eta^{\prime\prime}(E_x^*E_y + E_xE_y^*)\mathbf{L} \label{injection current, SI}\\
    \mathbf{J}_{S} = &\sigma(E_x^*E_y - E_xE_y^*)\mathbf{L} \label{Shift current, SI}
\end{align}
where \(\mathbf{J}_{I}\) and \(\mathbf{J}_{S}\) denote the optically excited injection and shift currents, respectively. The coefficients \(\eta^{\prime} = \frac{1}{2}(\eta^{xxxy} + \eta^{xyyy})\) and \(\eta^{\prime\prime} = \mathrm{Re}(\eta_{xxyx})\) represent the symmetry-allowed second-order optical conductivities for the injection current, while \(\sigma = i\,\mathrm{Im}(\sigma_{xxyx})\) is the symmetry-allowed counterpart for the shift current. The vector \(\hat{n}\) denotes the direction normal to the monolayer plane. Equation~\eqref{Shift current, SI} shows that the shift current always propagates parallel to \(\mathbf{L}\). In contrast, Eq.~\eqref{injection current, SI} reveals that the injection current comprises two components: one that always propagates perpendicular to \(\mathbf{L}\), and another that depends on the relative orientation between \(\mathbf{L}\) and the pump polarization. Equations~\eqref{injection current, SI} and~\eqref{Shift current, SI} not only identify the current flow directions, but also provide a simple method to determine the direction of \(\mathbf{L}\) in reverse.

In the main text, we use a quarter-wave plate (QWP) to modulate the pump polarization while consistently measuring the \(x\)-component of the emitted THz signal. Under this configuration, Eq.~\ref{Asymmetric Current} becomes:
\begin{align}
    J_x^{\text{asym}} \propto (\eta^{\prime} + \tfrac{1}{2}\eta^{\prime\prime})\cos{\theta_L} + \tfrac{1}{2}\eta^{\prime\prime}\cos{(4\phi - \theta_L)} + \sigma \sin{\theta_L}\sin{2\phi} \label{QWP, SI}
\end{align}
Comparing with Eq.~(3) in the main text, we identify the terms as \(L_1 \propto (\eta^{\prime} + \tfrac{1}{2}\eta^{\prime\prime})\cos\theta_L\), \(L_2 \propto \tfrac{1}{2}\eta^{\prime\prime}\), and \(C \propto \sigma \sin{\theta_L}\).

Additionally, we employ a half-wave plate (HWP) to linearly rotate the pump polarization. In this case, Eq.~\ref{Asymmetric Current} becomes:
\begin{equation}
    E_x(t) \propto \eta^{\prime}\cos{\theta_L} + \eta^{\prime\prime}\cos{(2\theta - \theta_L)}  \label{HWP, SI}
\end{equation}\\

\begin{table}[ht]
\centering
\begin{minipage}[t]{0.49\linewidth}
\centering
\begin{tabular}{|c|c|c|c|c|c|c|}
\hline
$D_{3d}$ & $1$ & $3$ & $2_{1\bar{1}0}$ & $-1$ & $-3$ & $m_{1\bar{1}0}$ \\
\hline
$A_{1g}$ & $1$ & $1$ & $1$ & $1$ & $1$ & $1$ \\
\hline
$A_{2g}$ & $1$ & $1$ & $-1$ & $1$ & $1$ & $-1$ \\
\hline
$A_{1u}$ & $1$ & $1$ & $1$ & $-1$ & $-1$ & $-1$ \\
\hline
$A_{2u}$ & $1$ & $1$ & $-1$ & $-1$ & $-1$ & $1$ \\
\hline
$E_g$    & $2$ & $-1$ & $0$ & $2$ & $-1$ & $0$ \\
\hline
$E_u$    & $2$ & $-1$ & $0$ & $-2$ & $1$ & $0$ \\
\hline
\end{tabular}
\end{minipage}%
\hfill
\begin{minipage}[t]{0.49\linewidth}
\centering
\begin{tabular}{|c|c|}
\hline
\multicolumn{2}{|c|}{irreps} \\
\hline
$A_{1g}$ & $E_x E_x^* + E_y E_y^*$ \\
\hline
$A_{2g}$ & $E_x E_y^* - E_y E_x^*$ \\
\hline
$E_g$ & $(E_x E_x^* - E_y E_y^*,\; E_x E_y^* + E_y E_x^*)$ \\
\hline
$E_u$ & $(J_x, J_y),\; (-L_y, L_x)$ \\
\hline
\end{tabular}
\end{minipage}
\caption{Character table of the group $D_{3d}$ and irreps}\label{D3d_and_irrep_tables}
\end{table}

\noindent\textbf{Symmetry calculation with group invariants in magnetic point group}\\
In the AFM state, the in-plane \(\mathbf{L}\) may point in arbitrary directions due to weak anisotropy. As a result, the presence of \(\mathbf{L}\) can break the lattice mirror symmetry, seemingly contradicting the conclusion from the previous section that mirror symmetry can be applied to simplify the analysis in a monolayer. To resolve this discrepancy, we also derive the symmetry constraints based on the magnetic point group and compare them with our previous calculations.
 
Following the theoretical analysis in Ref.~\cite{RN177}, we consider two representative cases where $\mathbf{L}$ aligns along the lattice $x$-axis (AFM-$x$) or the $y$-axis (AFM-$y$). 
In the AFM-$x$ configuration, a mirror symmetry plane perpendicular to the $y$-axis is preserved, corresponding to the magnetic point group $2^\prime/m$. In contrast, the AFM-$y$ configuration retains a $C_{2y}$ rotation symmetry but no mirror symmetry, giving rise to the magnetic point group $2/m^\prime$. If $\mathbf{L}$ points along a general in-plane direction (neither $x$ nor $y$), the magnetic point group reduces further to $1^\prime$, indicating there is neither mirror symmetry nor in-plane rotation symmetry. 

We begin with the AFM-$x$ case. Since $2^\prime/m = C_s \otimes \{\mathcal{I}, \mathcal{PT}\}$, and the magnetic photocurrent is invariant under $\mathcal{PT}$, it suffices to analyze the problem using the irreps of the group $C_s$. This significantly simplifies the symmetry analysis. Based on calculations analogous to those in the previous section or in Ref.~\cite{RN177}, one finds that there are three independent invariants for the injection current and one for the shift current. Using the general form $j_i = \sigma_{ijk} E_j E_k^*$, we can identify the corresponding nonzero tensor components, yielding
\begin{align}
\begin{bmatrix}
    J_x \\
    J_y
\end{bmatrix}
=
\begin{bmatrix}
    \text{Re}(\sigma_{xxy})(E_x E_y^* + E_y E_x^*) \\
    \sigma_{yxx} E_x E_x^* + \sigma_{yyy} E_y E_y^*
\end{bmatrix}
+
\begin{bmatrix}
    \text{Im}(\sigma_{xxy})(E_x E_y^* - E_y E_x^*) \\
    0
\end{bmatrix} \label{magnetic point, rank3}
\end{align}

As before, we expand $\sigma_{ijk}(\mathbf{L}) = \sigma_{ijkl}L_l + \mathcal{O}(\mathbf{L}^3)$, where $\sigma_{ijkl}$ is constrained by the $C_3$ lattice rotation symmetry. 
Combining Eq.~\eqref{magnetic point, rank3} with the symmetry constraints in Eq.~\eqref{C3 Constraint}, we obtain the photocurrent for the AFM-$x$ case:
\begin{align}
\begin{bmatrix}
J_x \\
J_y
\end{bmatrix}
&= \frac12(\sigma_{xxxy} + \sigma_{xyyy})
    (|E_x|^2 + |E_y|^2)
    \begin{bmatrix}
    0 \\
    -L_x
    \end{bmatrix}
+ i\, \operatorname{Im}(\sigma_{xxyx})
    (E_x^* E_y - E_y^* E_x)
    \begin{bmatrix}
    L_x \\
    0
    \end{bmatrix} \notag\\
&\quad + \, \operatorname{Re}(\sigma_{xxyx})
    \begin{bmatrix}
    (E_x^* E_y + E_y^* E_x) L_x \\
    -(|E_x|^2 - |E_y|^2) L_x
    \end{bmatrix} \label{AFM-x}
\end{align}

For the AFM-$y$ configuration, a similar derivation gives:
\begin{align}
\begin{bmatrix}
J_x \\
J_y
\end{bmatrix}
&= \frac12(\sigma_{xxxy} + \sigma_{xyyy})
    (|E_x|^2 + |E_y|^2)
    \begin{bmatrix}
    L_y \\
    0
    \end{bmatrix}
+ i\, \operatorname{Im}(\sigma_{xxyx})
    (E_x^* E_y - E_y^* E_x)
    \begin{bmatrix}
    0 \\
    L_y
    \end{bmatrix} \notag\\
&\quad + \, \operatorname{Re}(\sigma_{xxyx})
    \begin{bmatrix}
    -(|E_x|^2 - |E_y|^2) L_y \\
    - (E_x^* E_y + E_y^* E_x) L_y
    \end{bmatrix} \label{AFM-y}
\end{align}

Equations~\eqref{AFM-x} and \eqref{AFM-y} represent special cases of Eq.~\eqref{Asymmetric Current}, corresponding to \(\mathbf{L} \parallel \hat{x}\) and \(\mathbf{L} \parallel \hat{y}\), respectively. By combining these two equations, we seemingly recover the full expression of Eq.~\eqref{Asymmetric Current}. However, there is a subtle but important distinction between these cases. In Eqs.~\eqref{AFM-x} and \eqref{AFM-y}, the Néel vector \(\mathbf{L}\) is restricted to lie strictly along the \(x\)- or \(y\)-axis. Although our expansion is only to first order in \(\mathbf{L}\), all higher-order contributions in these cases are also constrained: when \(\mathbf{L} \parallel \hat{x}\), mirror-asymmetric terms vanish, and when \(\mathbf{L} \parallel \hat{y}\), in-plane rotationally asymmetric terms vanish. In contrast, Eq.~\eqref{Asymmetric Current} allows for \(\mathbf{L}\) in an arbitrary in-plane direction, so mirror- or rotation-asymmetric terms can emerge. Nevertheless, such asymmetric contributions only appear at third order or higher in \(\mathbf{L}\) (i.e., \(\sim \mathbf{L}^3\)).

In conclusion:
\begin{itemize}
    \item In a monolayer system, to first order in \(\mathbf{L}\), the circularly induced shift current always flows parallel to \(\mathbf{L}\), while the injection current—excited by the total light intensity \((|E_x|^2 + |E_y|^2)\)—always flows perpendicular to \(\mathbf{L}\).
    \item Layer stacking breaks lattice mirror symmetry, allowing both the shift and injection currents to deviate from their ideal directions with respect to \(\mathbf{L}\). This effect, however, is expected to be small due to the weak interlayer coupling.
    \item The Néel vector \(\mathbf{L}\) itself can also break mirror symmetry. This leads to additional asymmetric terms in the photocurrent, which appear at third order or higher in \(\mathbf{L}\).

\end{itemize}


\noindent\textbf{Dzyaloshinskii–Moriya interaction in MnPSe\textsubscript{3}}

In this chapter, we use irreps to analyze the symmetry-allowed Dzyaloshinskii–Moriya interaction (DMI), following the method outlined in Ref.~\cite{dzyaloshinsky1958thermodynamic}. DMI is forbidden in systems with lattice inversion symmetry~\cite{moriya1960anisotropic}. However, at the surface, inversion symmetry is naturally broken, thereby allowing DMI to emerge. 

For a monolayer at the surface, the relevant point group is \(C_{3v}\). We define \(\mathbf{L} = \frac{1}{2}(\mathbf{S}_1 - \mathbf{S}_2)\) and \(\mathbf{m} = \frac{1}{2}(\mathbf{S}_1 + \mathbf{S}_2)\), where \(\mathbf{S}_i\) is the spin at lattice site \(i\). The character table~\cite{aroyo2011crystallography} and irreps of \(C_{3v}\) are summarized in Table~\ref{C3v_and_irrep_tables}.

\begin{table}[ht]
\centering
\begin{minipage}[t]{0.49\linewidth}
\centering
\begin{tabular}{|c|c|c|c|}
\hline
$C_{3v}$ & $1$ & $3$ & $m_{1\bar{1}0}$ \\
\hline
$A_{1}$ & $1$ & $1$ & $1$ \\
\hline
$A_{2}$ & $1$ & $1$ & $-1$ \\
\hline
$E$     & $2$ & $-1$ & $0$ \\
\hline
\end{tabular}
\end{minipage}%
\hfill
\begin{minipage}[t]{0.49\linewidth}
\centering
\begin{tabular}{|c|c|}
\hline
\multicolumn{2}{|c|}{Irreps} \\
\hline
$A_{1}$ & $L_z$ \\
\hline
$A_{2}$ & $m_z$ \\
\hline
$E$     & $(m_x, m_y),\; (-L_y, L_x)$ \\
\hline
\end{tabular}
\end{minipage}
\caption{Character table and irreducible representations of the point group $C_{3v}$.}
\label{C3v_and_irrep_tables}
\end{table}

From the table, we see that the term \(m_x L_y - m_y L_x\) transforms as the irrep \(A_1\), meaning it is invariant under all operations of the point group \(C_{3v}\). Consequently, a symmetry-allowed contribution to the free energy can be written as \(F = -D(m_x L_y - m_y L_x)\), where \(D\) is a coupling coefficient. This expression can be rewritten in vector form as \(F = -\mathbf{D} \cdot (\mathbf{m} \times \mathbf{L})\), where \(\mathbf{D} = (0, 0, D)\) is the DM vector.

This result implies that when a weak net magnetization \(\mathbf{m}\), induced by an external magnetic field, lies perpendicular to the Néel vector \(\mathbf{L}\), it couples to \(\mathbf{L}\) such that reversing \(\mathbf{m}\) leads to a reversal of \(\mathbf{L}\). In other words, the in-plane Néel vector \(\mathbf{L}\) is coupled to a weak perpendicular magnetization \(\mathbf{m}\). This mechanism naturally explains our experimental observation: field cooling with an in-plane magnetic field perpendicular to \(\mathbf{L}\) selects specific magnetic domains, whereas a field applied parallel to \(\mathbf{L}\) does not.



\end{document}